%% file: angelaromano_ckm2014.tex
\documentclass[12pt]{article}
\usepackage{graphicx}

\newcommand\pubnumber{SNSN-323-63}
\newcommand\pubdate{\today}

\def\bham{School of Physics and Astronomy\\
University of Birmingham, B15 2TT Edgbaston, Birmingham, United Kingdom}

\def\support{\footnote{On behalf of the NA62 Collaboration:
G.~Aglieri Rinella, R.~Aliberti, F.~Ambrosino, B.~Angelucci, A.~Antonelli, G.~Anzivino, 
R.~Arcidiacono, I.~Azhinenko, S.~Balev, J.~Bendotti
% ...
\\
        }}
\def\Title#1{\begin{center} {\Large #1 } \end{center}}
\def\Author#1{\begin{center}{ \sc #1} \end{center}}
\def\Address#1{\begin{center}{ \it #1} \end{center}}

\newcommand\pubblock{\rightline{\begin{tabular}{l} \pubnumber\\
         \pubdate  \end{tabular}}}
\newenvironment{Abstract}{\begin{quotation}  }{\end{quotation}}
\newenvironment{Presented}{\begin{quotation} \begin{center} 
             PRESENTED AT\end{center}\bigskip 
      \begin{center}\begin{large}}{\end{large}\end{center} \end{quotation}}

\input econfmacros.tex
\newcommand{\pnn}{$K^+ \rightarrow \pi^+ \nu \bar{\nu}$ }
\newcommand{\pnnn}{$K_L \rightarrow \pi^0 \nu \bar{\nu}$ }

\begin{document}
\begin{titlepage}
\pubblock

\vfill
\Title{The \pnn decay in the NA62 experiment at CERN}
\vfill
\Author{ Angela Romano\support}
\Address{\bham}
\vfill
\begin{Abstract}
The main aim of the NA62 experiment at CERN is to study the rare \pnn decay 
and measure its Branching Ratio (BR) with $10\%$ precision. 
Due to its theoretical precision, this decay is an excellent probe to test the presence of New Physics (NP) at the highest scale complementary to LHC. 
At less than one month from the starting of the NA62 pilot run, the motivations, strategy and status of the experiment are described.

\end{Abstract}
\vfill
\begin{Presented}
The 8th International Workshop on the CKM Unitarity Triangle (CKM 2014)\\
Vienna, Austria, September 8-12, 2014
\end{Presented}
\end{titlepage}
\def\thefootnote{\fnsymbol{footnote}}
\setcounter{footnote}{0}

\section{Introduction}

The \pnn decay is a Flavour Changing Neutral Current (FCNC) process
particularly interesting to study the physics of flavour:
in the Standard Model (SM) the decay is forbidden at tree level and highly CKM-suppressed;
its amplitude is dominated by short-distance dynamics and 
the rate can be computed with extremely high precision.
These properties make the \pnn decay very sensitive to NP and studies have shown 
how the combined measurements of BR(\pnn) and BR(\pnnn) can discriminate among several NP scenarios, 
including tree-level FCNC mediated by heavy gauge boson $Z'$, with sensitivity to $M(Z')$ beyond LHC~\cite{Buras}. 
The SM expectation is BR(\pnn) $= (7.81\pm0.75\pm0.29) \times 10^{-11}$ ~\cite{Brod}. 
The current experimental measurement, BR(\pnn) $=1.73^{+1.15}_{-1.05} \times 10^{-10}$~\cite{Artamonov}, is based on 7 candidates 
observed by the E787/E949 experiments at the Brookhaven National Laboratory (BNL). 

\section{Experimental Strategy}

The main goal of the NA62 experiment at CERN Super-Proton-Synchroton (SPS)~\cite{proposal} is to measure BR(\pnn) at $10\%$ precision. 
The strategy foresees the collection of $\sim$ 100 SM signal events in three years of data taking, starting in 2015. 
Considering the SM rate of the \pnn decay, this will be achieved with a high-intensity kaon flux 
($\sim 10^{13}$ $K^{+}$ decays in the fiducial volume) and a $10\%$ detector acceptance.
The final precision is reached by keeping the total systematic uncertainty and backgrounds below a certain level ($10\%$).
This will be obtained with a background rejection performed at a level of $10^{12}$ with detector redundancy
in order to ensure a better control of the background.
Unlike the experiments E787/E949, which used kaons at rest, the NA62 experiment
will be using high-momentum 75GeV$/c$ $K^+$ decaying in flight.
Due to the presence of two undetectable neutrinos in the final state, the \pnn decay signature is 
given by one positive-charged track in the final state identified as a pion
and matched to one positive-charged track in the beam, identified as a kaon, 
with no other particles detected. In the NA62 experiment, the most important
background processes to \pnn events are the main $K^+$ decay modes,
with branching ratios up to $10^{10}$ times greater than the one expected for the signal.
In Tab.~\ref{tab:kdecays} these decay modes are reported, together with the techniques intended to reject them.
In order to achieve a background rejection factor at a level of $10^{12}$,
a precise measurement of the event kinematics, hermetic photon vetoes and
particle identification are crucial for the success of the experiment.

\begin{table}[htb]
\begin{center}
\begin{tabular}{l|cc}  
Decay mode &  Branching ratio ($\%$) & Background rejection \\ \hline
$K^+\rightarrow \mu^+\nu_{\mu}$  &   63.55 $\pm$ 0.11     &     Kinematics + $\mu$-PID      \\
$K^+ \rightarrow \pi^+ \pi^0$ &  20.66 $\pm$ 0.08     &     Kinematics + $\gamma$-Veto     \\ 
$K^+\rightarrow \pi^+\pi^+\pi^-$  &   5.59 $\pm$ 0.04     &     Kinematics + $\pi^-$-Veto      \\
$K^+\rightarrow \pi^+\pi^0\pi^0$  &   1.76 $\pm$ 0.02     &     Kinematics + $\gamma$-Veto      \\
$K^+ \rightarrow \pi^0 e^+ \nu_e$ &  5.07 $\pm$ 0.04     &     e-PID + $\gamma$-Veto     \\ 
$K^+ \rightarrow \pi^0 \mu^+ \nu_{\mu}$ &  3.35 $\pm$ 0.03     &     $\mu$-PID + $\gamma$-Veto     \\
\hline
\end{tabular}
\caption{Dominant kaon decays, their BRs and corresponding rejection
techniques. 
}
\label{tab:kdecays}
\end{center}
\end{table}

\begin{figure}[htb]
\centering
\includegraphics[height=0.8in]{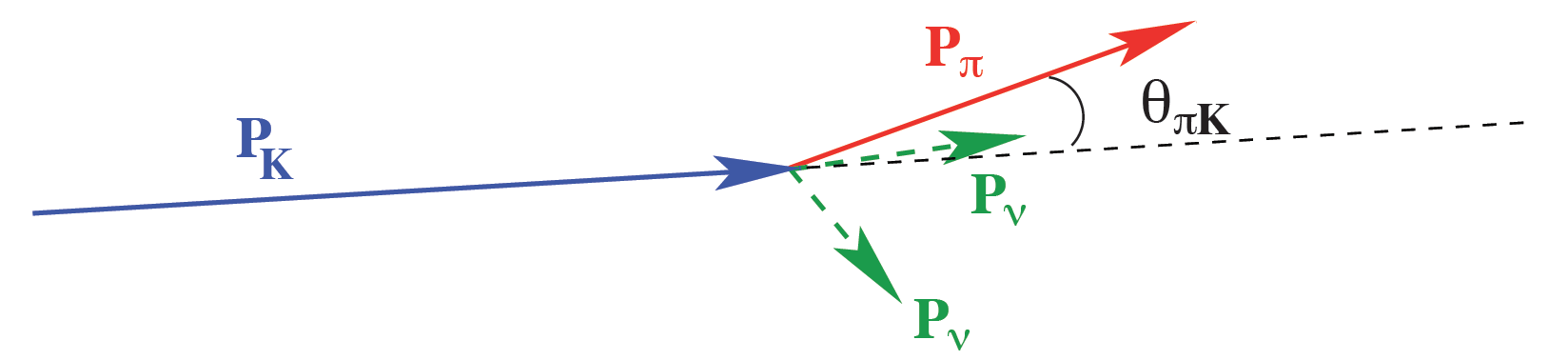}
\caption{Kinematics of the \pnn decay.}
\label{fig:kinematics}
\end{figure}

The kinematics of the \pnn decay is schematically sketched in Fig.~\ref{fig:kinematics}: 
the momentum of the incoming kaon $P_K$, the momentum of the outgoing pion $P_{\pi}$ 
and the angle between the mother and the daughter particles $\theta_{\pi K}$ are the only measurable quantities. 
The kinematics can be fully described by the squared missing mass variable $m^2_{miss} = (P_K -P_{\pi^+})^2$,
where $P_K$ is the kaon candidate 4-momentum and $P_{\pi^+}$ is the pion candidate 4-momentum. 


\begin{figure}[htb]
\centering
\includegraphics[height=2in]{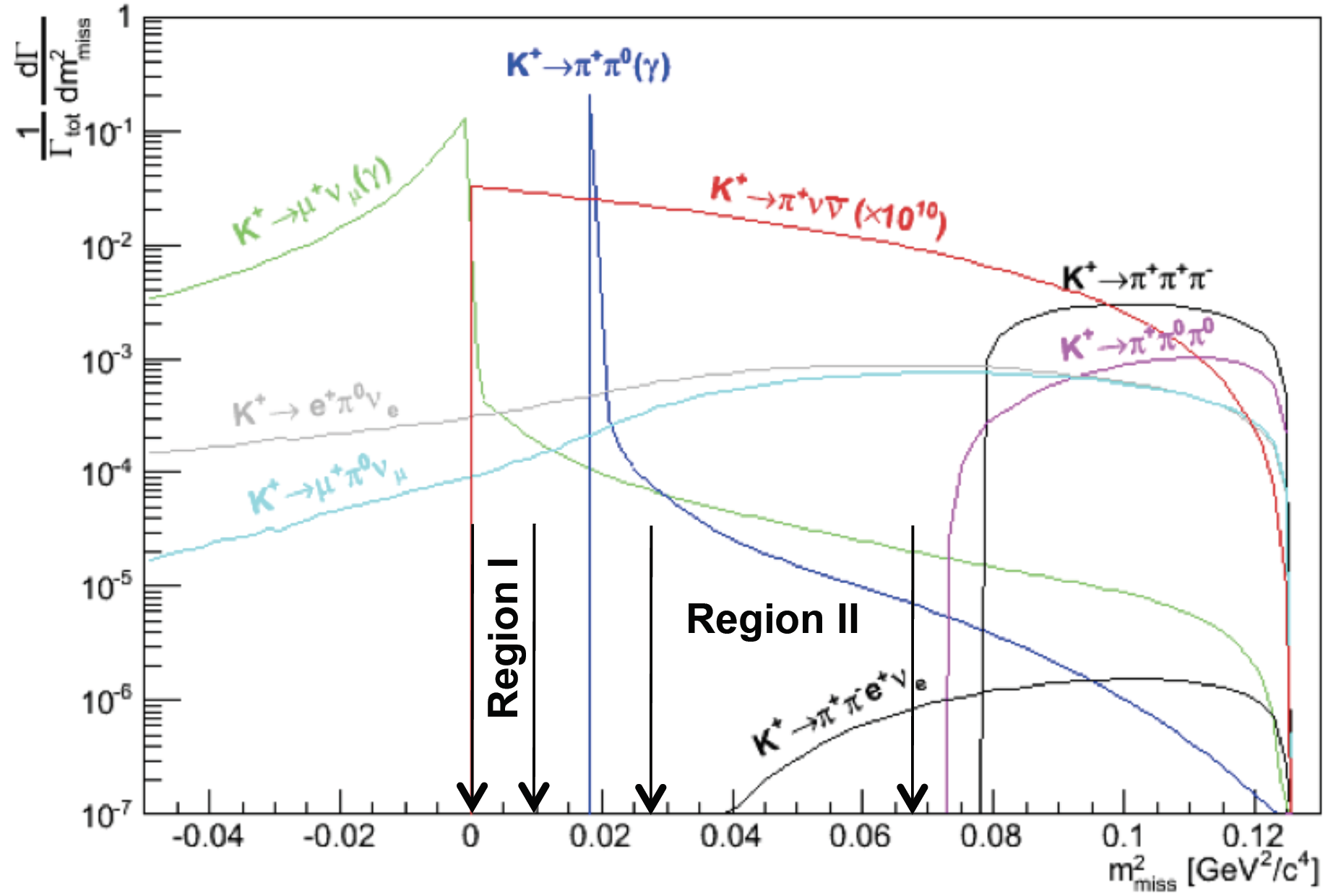}
\caption{Distributions of the $m^2_{miss}$ variable for signal and backgrounds from the main $K^+$ decay modes.
Backgrounds are normalised according to their BR; the signal is multiplied by a factor of $10^{10}$.
}
\label{fig:mmsqr}
\end{figure}

The distributions of $m^2_{miss}$ for signal and backgrounds from the main $K^+$ decay channels are shown in Fig.~\ref{fig:mmsqr}: the
backgrounds are normalised according to their branching ratio and the signal is multiplied by a factor $10^{10}$~\cite{ruggiero}.
The variable $m^2_{miss}$ can be used to reject $92\%$ of the backgrounds from the most frequent $K^+$ decay modes 
(first four channels in Tab.~\ref{tab:kdecays}).
Two region with minimum background can be defined: 
Region I above the $K^+\rightarrow \mu+\nu_{\mu}$ contribution and below the $K^+ \rightarrow \pi^+ \pi^0$ peak,
Region II above the $K^+ \rightarrow \pi^+ \pi^0$ contribution but below the $K \rightarrow 3\pi$ threshold (see Fig.~\ref{fig:mmsqr}).
Nevertheless, to guarantee the background rejection at the required level of $10^{12}$, 
kinematic constraints must be used in conjunction with PID and veto systems.

\section{The NA62 beam line and detector}

\begin{figure}[htb]
\centering
\includegraphics[height=2in]{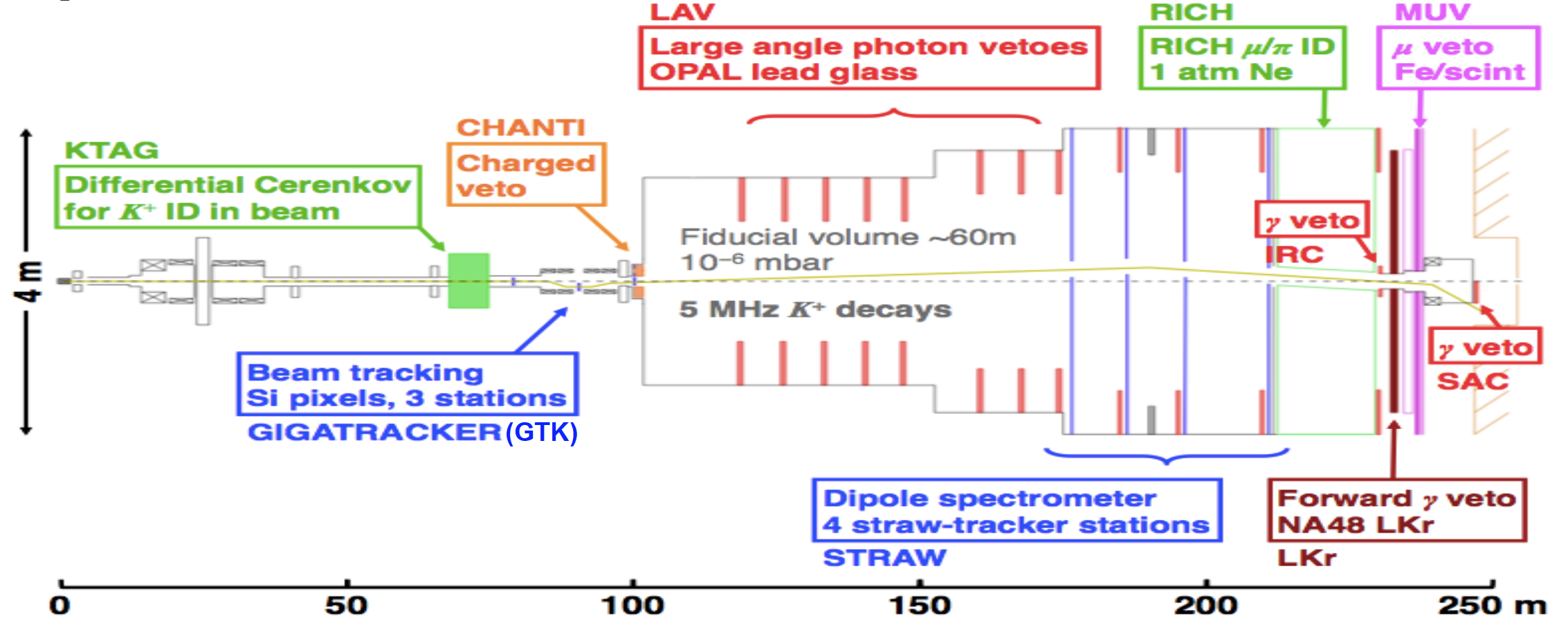}
\caption{Schematic layout of the NA62 experiment including the target, the beam line, the
decay fiducial region and detectors.}
\label{fig:na62_layout}
\end{figure}

NA62 will use a kaon decay-in-flight technique. 
A secondary unseparated hadron beam ($\sim 6\%$ kaons) of central momentum 75GeV$/c$ will be produced 
from SPS primary protons at 400GeV$/c$ directed on a beryllium target.
The number of $K^+$ decays integrated over a year in $65$m of decay fiducial length is $4.5\times10^{12}$; 
considering the SM signal BR(\pnn) and a $10\%$ detector acceptance, 
this translates in about $45$ signal events collected in one year of data taking.
The schematic layout of the NA62 experiment~\cite{TD} comprising the target, the beam line,
the decay fiducial region and detectors is shown in Fig.~\ref{fig:na62_layout}.
The decay region is contained in a large vacuum cylindrical tank kept at $10^{-6}$mbar pressure;
this allows to keep the background due to the beam scattering below the level of 1 event/year.

After the production target the first detector on the NA62 beam line is the KTAG,
an upgraded differential Cherenkov (CEDAR) counter, which is used to tag the $K^+$ in the beam.
Kaon identification is required to reduce at negligible level the background due to the beam
interaction with residual gas in the vacuum tank.
The KTAG identifies the $K^+$ component (45MHz rate) in the beam with respect to the other beam particles 
with at least $95\%$ efficiency and time resolution of the order of 100ps.
The next stage on the beam line is a particle tracking system composed by an achromatic corrector, 
made of four dipole magnets and three stations of Si micro-pixel GigaTracKer (GTK) detectors.
These stations are placed in vacuum and are crossed by the full beam intensity ($\sim 750$MHz).
The GTK provides the momentum and direction measurements for all particles in the beam,
with a resolution $\sigma(p)/p = 0.2\%$ and $\sigma_{\theta} = 16\mu$rad.
Moreover, in order to associate the right beam particle to the decay vertex, the GTK provides a time resolution better than 150ps.
A magnetic spectrometer, composed of four straw-tube chambers (STRAW) and a large-aperture dipole magnet,
is responsible for the detection and tracking of the charged decay products inside the vacuum tank. 
The magnet provides a transverse momentum kick of 270MeV/c and the expected momentum resolution is $\sigma(p)/p = (0.32 \bigoplus 0.008 \cdot p)\%$.
After the spectrometer, a Ring Imaging CHerenkov (RICH) counter is used for the identification of pions from kaon decays 
and the separation $\pi/\mu$, necessary to reject the $K^+\rightarrow \mu^+\nu_{\mu}$ decays, in the pion momentum range (15-35)GeV$/c$. 
The RICH is a cylindrical vessel with the beam pipe passing in its center and filled with Neon (at 1 atm). 
The RICH is expected to provide a time resolution of the pion candidate below $100$ps and a factor of $10^2$ on the rejection 
of the background from the $K^+ \rightarrow \mu^+ \nu_\mu$ decay.
As a consequence of the selected pion momentum range, for kaon decays with $K^+$ at 75GeV$/c$, 
the decay products, other than the accepted $\pi^+$, will carry an energy of at least 40GeV,
which helps the background rejection. 
As an example, the probability that both photons from a 40GeV $\pi^0$ decay are left undetected because of detector inefficiencies is below $10^{-8}$. 
The vacuum tank, the magnetic spectrometer and the RICH are interspaced
by twelve stations of ring-shaped Large Angle photon Veto (LAV) detectors. The LAV
is designed to reject the second main background due to $K^+ \rightarrow \pi^+ \pi^0$ decays, followed
by $\pi^0 \rightarrow \gamma\gamma$; together with the LKr electromagnetic calorimeter, which is inherited
from the NA48 experiment~\cite{Fanti}, it covers the angular region
of photon emission up to 50mrad. In the forward region the detection of photons
is rendered hermetic by means of an Intermediate Ring Calorimeter (IRC) and a Small-Angle Calorimeter (SAC). 
The LAV, LKr, IRC and SAC detectors compose the photon veto system.
The LAV stations are made of rings of lead-glass blocks recovered from the OPAL electromagnetic calorimeter barrel.
The LKr electromagnetic calorimeter is a quasi-homogeneous ionisation chamber with energy resolution 
$\sigma_E/E = 0.032/ \sqrt(E) \bigoplus 0.09/E \bigoplus 0.0042$ (E in GeV).
The SAC and IRC are two shashlik calorimeters each made of 70 iron-scintillator planes.
The muon veto system (MUV) provides a factor of $10^{5}$ rejection of muons from the most frequent kaon
decay $K^+ \rightarrow \mu^+ \nu_\mu$ . It is composed of three sub-detectors, called MUV1, MUV2, and MUV3.
The first two modules follow directly the LKr calorimeter and work as hadronic calorimeters. 
Both modules are classic iron-scintillator sandwich calorimeters. The MUV3 detector, located after a
80cm thick iron wall, is made of scintillator tiles. It serves as fast muon veto ($\sigma_T \sim 1$ns).
At the end of the NA62 beam line, the particles are finally absorbed in a beam dump composed of iron surrounded by concrete.

The intense flux needed in the NA62 experiment implies the design of high performance triggering and data acquisition systems, 
which minimise the dead time while maximising data collection reliability.
The particle rate in the detectors following the decay region is expected to be $\sim10$MHz.
The trigger hierarchy is made of three logical levels:
a hardware L0 trigger, based on the input from a few sub-detectors;
a software L1 trigger, based on information computed independently by each sub-detector system;
a software L2 trigger, based on assembled and partially reconstructed events.
The L0 trigger reduces the rate to $\sim 1$MHz, while the final rate output from the L2 trigger is $\sim$ 10kHz.

\vspace{-0.1cm}
\section{Conclusions}
The NA62 detector installation has been completed in September 2014 and the start of a two-month pilot run is scheduled on the 6th October 2014.
The main goals in 2014 are: the commissioning of the hardware and readout with particles at low intensity and the addressing of the \pnn SM sensitivity.
The main NA62 data taking at the nominal intensity is expected in years 2015-2017, when the collection of about 100 SM signal events 
in three years of data taking for the measurement of BR(\pnn) at $10\%$ precision is foreseen. 



\end{document}

%% file: econfmacros.tex
\def\beq{\begin{equation}}
\def\eeq#1{\label{#1}\end{equation}}
\def\eeqn{\end{equation}}

\def\beqa{\begin{eqnarray}}
\def\eeqa#1{\label{#1}\end{eqnarray}}
\def\eeqan{\end{eqnarray}}

\let\bar=\overbar

\def\Dslash{\not{\hbox{\kern-4pt $D$}}}
\def\dslash{\not{\hbox{\kern-2pt $\del$}}}

\def\msb{{\bar{\ssstyle M \kern -1pt S}}}

